\documentclass[11pt]{article}
\usepackage{latexsym}
\raggedright
\newcommand{\T}{{\rm Tr} \,}

\newcommand{\cB}{{\cal B}}
\newcommand{\cH}{{\cal H}}

\newcommand{\cL}{{\cal L}}

\newcommand{\1}{{\bf 1}}

\newcommand{\rc}{{\rm anti}}
\begin{document}
  \title{Quantum information transfer\\
         from one system to another one}
  \author{Armin Uhlmann}
  \date{Institute
  for Theoretical Physics\\ University of Leipzig}
  \maketitle

\begin{abstract}
The topics of the paper are: a) Some anti-linear maps governing
EPR tasks if no reference bases are distinguished. b)  Imperfect
quantum teleportation and the composition rule. The ancilla
is supposed pure but otherwise arbitrary. c) Quantum
teleportation with distributed measurements. d) Remarks
on EPR with a mixed state, triggered by a
L\"uders   measurement.
\end{abstract}

\section{Introduction}
The problem of transferring ``quantum information'' from one
quantum system into another one has its roots in the 1935
paper \cite{EPR35} of A.~Einstein, B.~Podolski, and
N.~Rosen. These authors posed a far reaching question, but
they doubt the answer given by quantum theory. The latter,
as was pointed out by them,
asserts the possibility to create simultaneously and
at different places exactly the same random events.
The phenomena is often called ``EPR effect'' or, simply,
``EPR''.

Early contributions to the EPR problem are due to
Schr\"odinger, \cite{Schr35a}.
Since then a wealth of papers had appeared on the subject,
see \cite{Peres93} and \cite{NC00} for a r\'esum\'e.
Even to-day some authors consider it more a ``paradox''
than a physical ``effect'',
because EPR touches the question,
whether and how space and time can live with the very
axioms of quantum physics, axioms which, possibly, are prior to
space and time\footnote{Sometimes it seems helpful to think
space-time a user interface of the quantum world.}.

Quantum information theory considers EPR  as a map or as
a ``channel'', as an element of
protocols transferring ``quantum information'' from one
system to another one or
supporting the transmission of classical information,
\cite{Ekert91}, \cite{BeWi92}, \cite{NC00}. One of our
aims is to present a certain calculus for EPR and
EPR-like processes or, more general, for processes triggered
by measurements. We begin, therefore, with some
selected fundamentals of quantum measurements.

The following
treatment of EPR has its origin in the identification
problem in comparing two or more quantum systems. It is
by far not obvious how to identify two density operators, say
$\varrho^a$ and $\varrho^b$, belonging to two different
Hilbert spaces, $\cH_a$ and $\cH_b$. Often one fixes two
bases, $\{ \phi^a_j \}$ and $\{ \phi^b_j \}$, and defines
$\varrho^a$ and $\varrho^b$ to be ``equal'' one to another
if they have the same matrix representation with respect
to the reference bases. The more, one needs a stable
synchronization if several tasks have to be done in the
course of time. It seems, therefore, worthwhile to
postpone the selection of the reference bases as long as
possible. If that can be done, it can be done using
the s-maps, \cite{Uh00a}, of the EPR section.
These maps are
anti-linear. The anti-linearity in the EPR problem
is usually masked by the reference bases: The bases provide
conjugations which create, combined with the ``natural''
anti-linearity, the suggestion of an unrestricted
linearity. An interesting, though quite different
approach, \cite{BO00}, is founded on
Ohya's idea of compound states \cite{Oh83}. Also in
\cite{Fivel95} there is a side remark on anti-linearity.

For pure states in quantum systems with finitely many degrees
of freedom, there is a duality between pure states and
maximal properties in the sense of von Neumann and Birkhoff.
In the section ``inverse'' EPR we show by an example
the meaning of the mentioned duality.

We proceed with the beautiful quantum teleportation protocol
of Bennett at al \cite{BBCJPW93}.
Here we prove a composition rule for imperfect (i.~e.~not faithful)
quantum teleportation.
Then we show its use in quantum teleportation
with distributed measurements by an example with a 5-partite
system, and an EPR example based on a 4-partite Hilbert space.

There is a short section on polar decompositions of the
s-maps, including a quite elementary link to operator
representations. From a physical point of view, $^*$-representation
theory provides a classification of the ways a quantum system
can be embedded in a larger one. However, this topic is outside
the realm of the present paper.

Finally we show how to handle, again by some anti-linear
maps, an EPR task in a bi-partite system if its state is mixed
and if a measurement is performed in one of its subsystems by
a projection operator of any rank.

\underline{Remarks on notation:} In this paper the Hermitian adjoint of a map
or  of an operator $A$ is denoted by $A^*$. The scalar product
in Hilbert spaces is assumed linear in its second argument.

\section{Preliminaries}
The implementation independence in quantum information theory
is guarantied by the use of Hilbert spaces, states
(density operators), and operations between and on them. It
is {\it not} said,
what they physically describe in more concrete terms,
whether we are dealing with spins, polarizations,
energy levels, particle numbers, or whatever you can imagine.
Because of this, the elements of quantum information theory,
to which the EPR-effect belong, are of rather abstract nature.
\medskip

Let a physical system be is described by an Hilbert
space $\cH$. A quantum state of the system is then given
by a density operator $\omega$, a positive operator with
finite trace, the latter normalized to be one. Thus every
positive trace-class operator different from the zero operator
uniquely defines a state. One only has to divide it by its
trace.
(If the Hilbert space is not of finite dimension, there are
also so-called singular states. For our purposes we can
safely ignore them.)

Every vector $\psi \in \cH$, $\psi \neq 0$,
defines a vector state, the density operator of which is
the projection operator, say $P_{\psi}$, onto the 1-dimensional
subspace generated by $\psi$. It is common use to speak of
``the state $\psi$'' if the state can be described by
$P_{\psi}$.  The vector states of our system are called
{\em pure} if the properties of linear independent vectors
do not coincide.

The quantum version of Boolean Logics is due to
Birckhoff and J.~von Neumann, \cite{BvN36}.
According to them a {\em property,}
 a quantum state can have, is a subspace of $\cH$ or,
equivalently, a projection operator onto that subspace.
Not every subspace may be considered a property.
The point is, that there
are no other properties a quantum state can have. This
well established postulate excludes some hidden parameter
dreams.

 {\em Here we shall assume that every subspace defines
a property, and that two different subspaces encode
different properties.} It is another way to express
the purity assumption for vector states.

Looking at these two concepts, states and properties,
there is a certain ``degeneracy''. A vector can denote
a state or a property.
A (properly) minimal projection
operator represents {\em either} a maximal
property {\em or} the density operator of a
pure state. What applies depends on the context.
The existence of maximal
properties is a special feature of physical systems
with a finite number of degrees of freedom.

Let $\cH_0$ be a subspace of $\cH$ denoting a property.
A state, given by a density operator $\omega$, possesses
property $\cH_0$, if and only if its support is in
$\cH_0$. That is, $\omega$ must annihilate the orthogonal
complement of $\cH_0$. If $\omega = P_{\psi}$ is a
vector state, this is equivalent to $\psi \in \cH_0$.

Let $P_0$ denote the ortho-projection onto $\cH_0$.
A test, whether $\omega$ has property $P_0$ results
in one bit of information: Either the answer is YES
or it is NO.\\
i) The probability of outputting
the answer YES is $p := \T \, P_0 \omega$.\\
ii)  If $p$ is not zero, and if
the answer is YES, then
the test has {\em prepared} the new state
$P_0 \omega P_0$. Multiplying by $p^{-1}$
gives its density operator.\\
\medskip

An {\em executable measurement within $\cH$} is
characterized by a finite orthogonal decomposition
of $\cH$ into subspaces. The subspaces are assumed
to be properties. Denoting by $P_j$ the orthogonal
projections onto the subspaces, the requirement reads
\begin{equation} \label{o1}
\sum_{j=1}^m P_j = \1, \quad P_i P_k = 0 \, \hbox{ if }
\, i \neq k \, .
\end{equation}
\underline{Remark:} \, The
phrase ``executable'' asserts the possible existence
of an apparatus doing the measurement. A general observable
can be approximated (weakly) by such devises.
Important physical quantities like energy, momentum, and
position in Schr\"odinger theory represent examples of
observables, which can be approximated by executable ones
without being executable themselves.

\noindent \underline{Remark:} \, In saying that the
measurement is
``in'' $\cH$ we exclude measurements in an upper-system
containing the system in question as a sub-system. Such
a larger system allows for properties not present in the
smaller one. In allowing such measurements we arrive at the
so-called POVMs, ``positive operator valued measurements''.
\medskip

To be a measurement, the device testing the
properties $P_j$ should output a definite signal
$a_j$ if it decides to prepare the state $P_j \omega P_j$.
Well, $a_1, \dots, a_m$ constitute the letters of an
alphabet. The device randomly decides what letter to choose.
The probability of a decision in favor of the letter $a_j$
is $\T P_j \omega$ with $\omega$ the density operator
of the system's state. Thus, the classical information
per probing the properties (\ref{o1}) is
$$
H(p_1, \dots, p_n) = - \sum p_j \log_2 p_j ,
\quad p_j = \T \, P_j \omega \, .
$$
A little more physics come into the game in assuming
that the alphabet consists of m different complex numbers.
Then the operator
\begin{equation} \label{o2}
A := \sum_1^n a_j P_j
\end{equation}
is an {\em observable} for the measurement of the
properties (\ref{o1}). Clearly, the executable observables
are normal operators, $A A^* = A^* A$, and their spectra
are finite sets.

One observes that information theory is not interested in
the nature of the alphabet that distinguishes the outcomes
of a measurement. It suffices for its purposes to
discriminate the outcomes and to know the state that is
prepared. {\em Portability} is gained that way.
\medskip

It is standard that two properties can be checked
simultaneously if and only if their ortho-projections
commute. Otherwise one gets in trouble with the
probability interpretation.
Two observables, $A$ and $B$, can be measured
(or approximated by such procedures) simultaneously
provided they commute.
Executing a set $A_1, \dots, A_n$ of mutually commuting
observables will be called a {\em distributed measurement.}

Non-relativistically a distributed measurement may consist
of several measuring devices, sitting on different
(possibly overlapping) places
in space, but being triggered at the same time.

Relativistically,
every measurement is done in a certain space-time
or ``world'' region. A particular case of a
distributed measurement consists of devices
doing their jobs in disjunct, mutually space-like
world regions: Quantum theory does not enforce
restrictions for measurements (or ``interventions''
a la A.~Peres) for
space-like separated world regions. EPR and
quantum teleportation make use of it in an
ingenious way.

Thinking in terms
of the evolution of states in the course of time, these
tasks update the initial conditions of the
evolution. The choice of the  new Cauchy data
is done randomly and
governed by transition probabilities.

In Minkowski space the problem is somehow delicate.
According to
Hellwig and Kraus \cite{HK70} it is consistent to let
take place the state change at the boundary of the past of
the region. The past of the world region is the union
of all backward light-cones terminating in one of
the world points of the region the measurement is done.
Finkelstein \cite{Fi00} has argued that it is
also possible to allow the change at the light-like
future of the world region in question. We, \cite{CU},
think it even consistent to assume a slightly stronger
rule: {\em The state changes accompanied by a measurement
in a space-time region takes place at the set of
those points, which are neither in the past nor in the
future of that region.} The assumed region of influence
is bounded to the past a la Hellwig and Kraus and to
the future according to Finkelstein. The remarkable
experiments of Zbinden et al \cite{ZBGT00} agree with it.

Finally, I mention
some specialties in testing the properties of
vector states: If $\psi$ is a vector state, and $A$
an observable (\ref{o2}), any
vector, prepared by testing $A$, is of the form
$P_i \psi$. It follows that
the relative phase between $\psi$
and a non-zero $P_i \psi$ is
real and positive. Hence, the Hilbert space distance
between them equals their Fubini-Study distance:
The state changes by
 measurements proceeds along Study-Fubini geodesic
arcs.

Similar considerations with general
(``mixed'') states are more involved. These
states allow for quite different ``purifications'',
i.~e. lifts to vector states living
in larger quantum systems: One only gets
inequalities for the distance. However, the case of the
minimal possible distance is a distinguished one.

\section{EPR}
Let us consider a bi-partite quantum system composed
of two Hilbert spaces $\cH_a$ and $\cH_b$ and one of its
vectors
\begin{equation} \label{e1}
\cH := \cH_a \otimes \cH_b, \quad \psi \in \cH \, .
\end{equation}
In such a bi-partite system $\cH_a$ characterizes a
subsystem, the a-system, which is embedded in the
system of the
Hilbert space $\cH$. The same is with the b-system.

We assume the state of the composed system is the vector state
defined by $\psi$. We are interested in what is happening if
a property is checked in the a-system.
A {\em local subspace} of $\cH$ is a direct product of
two subspaces, one of $\cH_a$, the other one of $\cH_b$.
A {\em local property} of $\cH$ is, therefore, a
projection operator of the form $P_a \otimes P_b$.
$P_a$ and $P_b$ are projectors from the subsystems.
Similarly one proceeds in multi-partite systems.

If $P_a$ is a property of $\cH_a$, the local property
in the composed system that checks nothing in the
b-system reads $P_a \otimes \1_b$. If so, and if the
test of $P_a$ outputs YES, the newly prepared state
is again a vector state. The state change is
\begin{equation} \label{e2}
\psi \, \mapsto \, (P_a \otimes \1_b) \psi \, .
\end{equation}
Is something to be seen in the b-system by such a change?
Posing and answering the question is an essentially part
of the EPR problem. In pointing out the intrinsic
anti-linearity in the EPR problem we follow \cite{Uh00a}
and \cite{Uh00b}.

Let us consider maximal properties of the a-system,
\begin{equation} \label{e3}
P_a = { |\phi^a \rangle\langle \phi^a| \over
\langle \phi^a, \phi^a \rangle }, \quad \phi^a \in \cH_a
\, .
\end{equation}
Then the state prepared in (\ref{e2}) must
 be a product vector, the first
factor being a multiple of $\phi^a$. Therefore, given
$\psi$, there {\em must} be a map from $\cH_a$ into $\cH_b$
associating to any given $\phi^a$ its partner in the
product state. Let us denote this map by
$$
\cH_a  \ni \phi^a \, \mapsto \,
{\bf s}^{ba}_{\psi} \phi^a \in \cH_b \, .
$$
It is defined by
\begin{equation} \label{e4}
\bigl(|\phi^a \rangle\langle \phi^a| \otimes \1_b \bigr)
\, \psi = \phi^a \otimes {\bf s}^{ba}_{\psi} \phi^a, \quad
\forall \phi^a \in \cH_a \, .
\end{equation}
We see: {\em If in testing the property $\phi^a$
the answer is YES, the same is true with certainty if
in the b-system one is asking for the property
${\bf s}^{ba}_{\psi} \phi^a$.}

It becomes clear by inspection of (\ref{e4}) that ${\bf s}^{ba}_{\psi}$
is an {\em anti-linear} map from $\cH_a$ into $\cH_b$
which depends linearly on $\psi$.
We also may ask the same question starting from the b-system,
resulting in an anti-linear map ${\bf s}^{ab}_{\psi}$
from $\cH_b$ into $\cH_a$,
\begin{equation} \label{e5}
\bigl(\1_a \otimes |\phi^b \rangle\langle \phi^b| \bigr)
\, \psi = {\bf s}^{ab}_{\psi} \phi^b \otimes \phi^b, \quad
\forall \phi^b \in \cH_b \, .
\end{equation}
Let us go back to (\ref{e4}) and let us choose a vector
$\phi^b$ in $\cH_b$. Taking the scalar product (\ref{e4})
with $\phi^a \otimes \phi^b$, one easily finds
$$
\langle \phi^a \otimes \phi^b, \psi \rangle =
\langle \phi^b, {\bf s}^{ba}_{\psi} \phi^a \rangle .
$$
By symmetry, or by using (\ref{e5}) appropriately, one
finally arrives at the identity
\begin{equation} \label{e6}
\langle \phi^a \otimes \phi^b, \psi \rangle =
\langle \phi^b, {\bf s}^{ba}_{\psi} \phi^a \rangle =
\langle \phi^a, {\bf s}^{ab}_{\psi} \phi^b \rangle
\end{equation}
which is valid for all $\phi^a \in \cH_a$
and $\phi^b \in \cH_b$. Obviously, taking into account
their anti-linearity, the two maps between the Hilbert
spaces of the subsystems are Hermitian adjoints one from
another.
$$
({\bf s}^{ab}_{\psi})^* = {\bf s}^{ba}_{\psi},
\quad
({\bf s}^{ba}_{\psi})^* = {\bf s}^{ab}_{\psi}
$$
Finally, by the linearity of the s-maps with respect to
$\psi \in \cH$, one arrives at the following recipe
for their construction:
\begin{equation} \label{e7}
\psi = \sum a_{jk} \phi^a_j \otimes \phi^b_k \, \,
\Rightarrow \, \,
{\bf s}^{ba}_{\psi} \phi^a =
\sum a_{jk} \langle \phi^a, \phi^a_j \rangle \, \phi^b_k
\end{equation}
The exchange $a \leftrightarrow b$ of the letters a and b
in (\ref{e7}) produces the adjoint s-map.

The s-maps obey some simple rules if local operations are
applied to them. The most obvious is
\begin{equation} \label{e7a}
\varphi = (A \otimes B) \psi \, \leftrightarrow \,
{\bf s}^{ab}_{\varphi} = A \, {\bf s}^{ab}_{\psi} \, B^* \, .
\end{equation}
\medskip

Let us now escape from the formalism to a short discussion.
We assume, as starting point, the bi-partite system in a
pure state $\psi \in \cH$. We can assume that $\psi$
and an arbitrarily chosen $\phi^a$ are unit vectors. $P_a$
denotes the projection operator of the 1-dimensional subspace
generated in $\cH_a$ by $\phi^a$.

What can be seen from $\psi$ in the subsystems? This is
encoded in the reduced states, in the density operators
$\varrho^a_{\psi}$ and $\varrho^b_{\psi}$ respectively.
In more general terms: The state of
a subsystem is given by the expectation values of the
operators accessible within the subsystem.
All what an owner, say Bob, can learn within his subsystem
$\cH_b$ {\em without} resources from outside,
he has to learn from $\varrho^b$. Any
belief, he could learn anything else from its quantum
system alone, is nothing than a reanimation of the hidden
parameter story.

The reduced density operators can be
calculated by partial traces.
In the case at hand
a definition for the b-system is
$$
\langle \psi, (\1_a \otimes B) \psi \rangle = \T \,
\varrho^b_{\psi} B , \quad \forall \, B \in \cB(\cH_b) \, .
$$
The reduced density operators can also be expressed
by the s-maps,
\begin{equation} \label{e8}
\varrho^a_{\psi} = {\bf s}^{ab}_{\psi} {\bf s}^{ba}_{\psi},
\quad
\varrho^b_{\psi} = {\bf s}^{ba}_{\psi} {\bf s}^{ab}_{\psi}
\end{equation}
The probability, $p$, for a successful test of $\phi^a$ is
$\langle \phi^a, \varrho^a_{\psi} \phi^a \rangle$.
The maximal possible probability appears if $\phi^a$
is an eigenvector to the largest eigenvalue
of $\varrho^a_{\psi}$.

The square roots of the eigenvalues $p_i > 0$
of $\varrho^a_{\psi}$ are the Schmidt-coefficients of
the Schmidt decomposition of $\psi$ and,
according to (\ref{e8}), also
the singular values of ${\bf s}^{ab}_{\psi}$.

Let $\{ \phi_j^a \}$ be the vectors of a basis and $P_a^j$
the ortho-projection onto the space generated by
$\phi_j^a$.
Let us now ask what is going on if we test the
properties $P_a^j$.
We can use any operator
$$
A = \sum a_j P_a^j
$$
with mutually different numbers $a_j$.
The probability $p_j'$ of preparing
$\phi_j^a$ is
$\langle \phi_a^j, \varrho_{\psi}^a \phi_a^j \rangle$.
It is well known, that the probability vector $\{ p_j'\}$
is majorized by the set of eigenvalues $\{ p_j\}$
of $\varrho^a$. Any probability vector, which is majorized
by the vector of its eigenvalues, can be gained this
way by the use of  a suitable basis of $\cH_a$.
Consequently, in measuring $A$, one can produce a message
with an entropy not less than the entropy
of the eigenvalue distribution of $\varrho^a_{\psi}$.
If and only if the chosen basis is an eigen-basis of
$\varrho^a_{\psi}$, we get the minimally possible entropy.

Enhancing the entropy of Alice's side is not useful for Bob.
Though his system will definitely be in the state
$\phi_j^b = {\bf s}^{ba} \phi_j^a$ if on
Alice's side the state $\phi_j^a$ is prepared,
 he cannot always make too much use of it.
While Alice is preparing states which {\em must} be mutually
orthogonal, and hence distinguishable,
the vector states on Bob's side do not share this
 necessarily. Indeed, Bob's state are mutually orthogonal
if and only if Alice had minimized the entropy, i.~e.~if
she had chosen an eigen-basis of her density operator.

Let us repeat it from another perspective. Let Alice
perform some measurements using the observable $A$.
Assume that just before any measurement, the state
of the bi-partite system is the vector state $\psi$.
Then, whenever the device answers ``$a_j$'', the
state of the a-system changes to $\phi_j^a$. The state of
the b-system becomes $\phi_j^b = {\bf s}^{ba} \phi_j^a$.
Bob {\em knows} this state iff he knows $\psi$ and
which of the values $a_j$ the measuring device
has given to Alice.
Now, if Alice
uses an eigen-basis of the density operator $\varrho^a_{\psi}$
then Bob himself is able to measure which state he get
and, therefore, which $a_j$
Alice has obtained. On the contrary, if Alice does not
use a basis of eigenvectors, Bob's possible states are not
orthogonal and he cannot distinguish exactly
between them.
Therefore, the gain in entropy in the a-system by
using a measurement basis distinct from the
eigenvector basis is compensated
by a loss of Bob's possibility to distinguish between
the states he gets.

One can prove the assertion by calculating
$$
\langle \phi_j^b, \phi_k^b \rangle =
\langle {\bf s}^{ba} \phi_j^a, {\bf s}^{ba} \phi_k^a \rangle
=
\langle  \phi_k^a, {\bf s}^{ab} {\bf s}^{ba} \phi_j^a \rangle
$$
or, by (\ref{e8}),
\begin{equation} \label{e9}
\langle \phi_j^b, \phi_k^b \rangle =
\langle  \phi_k^a,  \varrho^a_{\psi} \phi_j^a \rangle \, .
\end{equation}
If {\em all} von Neumann measurements of  Alice are on
equal footing, and Bob can {\em always} discover
the state prepared by Alice within his system to
any precision, the EPR settings is ``perfect'' or ``tight''.
In the tight case the reduced
density operator $\varrho^a_{\psi}$ of Alice is equal to
$(\dim \cH_a)^{-1} \1_a$, i.e.~to the unique tracial state of
her system. This state is like  ``white quantum paper'',
there is no quantum information at all in it. The
``more white'' Alice's ``quantum paper'' $\varrho^a_{\psi}$ is,
the better EPR is working. That somewhat fabulous language
can be made precise substituting ``more mixed'' or
``less pure'' for ``more white''.

A further remark should be added to our short and incomplete
account of the EPR mechanism. It is a well known theorem
that $\cH_a \otimes \cH_b$ is canonically isomorph to
the space $\cL^2(\cH_a, \cH_b^*)$  of Hilbert-Schmidt
mappings form $\cH_a$ into $\cH_b^*$. On the other
hand, $\cH_b^*$ is canonically anti-linearly isomorphic
to $\cH_b$, a fact used by P.~A.~Dirac to establish his
bra-ket correspondence $|.\rangle \leftrightarrow \langle .|$
Composing both maps one immediately see the isomorphism
between $\cH_a \otimes \cH_b$ and the space of anti-linear
Hilbert-Schmidt maps $\cL^2(\cH_a, \cH_b)_{\rc}$. The
isomorphism is an isometry expressed by
\begin{equation} \label{e10}
\langle \varphi , \psi \rangle =
\T \, {\bf s}_{\psi}^{ab} {\bf s}_{\varphi}^{ba} =
\T \, {\bf s}_{\psi}^{ba} {\bf s}_{\varphi}^{ab}
\end{equation}
with $\psi$ and $\varphi$ from $\cH = \cH_a \otimes \cH_b$.

\section{``Inverse'' EPR}
In the preceding section we have considered three vectors:
$\psi$ from the composite Hilbert space (\ref{e1}) and $\phi^a$,
$\phi^b$ from its constituents. In the EPR setting $\psi$
is a given pure state which is to test whether it enjoys
the local properties defined either by $\phi^a$,
by $\phi^b$, or by both. In the ``dual'' or ``inverse''
EPR setting their roles are just reversed: $\psi$
appears as a non-local property which is to check.
$\phi^a \otimes \phi^b$ is the state to be tested for
the property $\psi$. Because transition probabilities
are symmetric in their arguments, one can
enrol the EPR setting backwards. The trick has been
clearly seen and used by
C.~Bennett, G.~Brassard, C.~Crepeau, R.~Jozsa, A.~Peres,
and W.~Wootters in their famous quantum teleportation
paper \cite{BBCJPW93}, see also \cite{Br98}.

To demonstrate what is going on, let us consider a simple
but instructive example. Here $\cH$ is of dimension four,
and its two factors 2-dimensional. Dirac's bra-ket notation
is used, but in scalar products 
anti-linear maps should be applied
to kets only! In the example we choose the vectors
$$
\psi = {1 \over \sqrt{2}} (|00\rangle + |11\rangle),
\quad \phi^a = |x\rangle, \, \, |\phi^b = |0\rangle
$$
with $x = 0, 1$. Alice is trying to send a bit-encoded
message to Bob by choosing $|x\rangle$ accordingly one
after the other. Bob's input is always $|0\rangle$.
By doing so, they enforce the bi-partite system into the state
$$
|x \, 0\rangle = |x\rangle \otimes |0\rangle, \quad x = 0, 1
$$
Then it is checked whether it has the property $\psi$.
\noindent If $x=1$, the measuring apparatus
will necessarily answer
the question with NO because the state is orthogonal to
$\psi$. If, however, $x=0$, the answer is YES with
probability $0.5$ and NO with the same probability. The
input state $|00\rangle$ of the bi-partite
system now has changed as follows:
$$
\hbox{YES} \, \mapsto \psi, \quad
 \hbox{NO} \, \mapsto \psi' =
{1 \over \sqrt{2}} (|00\rangle - |11\rangle) \, .
$$
Let now $q$ be the probability of an input $x=0$. Then
the input ensemble is transformed by the measurement
in the following way:
$$
\{ |00\rangle , |10\rangle ; q , 1-q \} \, \to
\, \{\psi , \psi' , |10\rangle ; {q \over 2} , {q \over 2} , 1-q \}
$$
The classical information encoded in the input state is not
lost. It could be regained by measuring the property
$$
|\psi \rangle\langle \psi| + |\psi' \rangle\langle \psi'| ,
$$
a task which does not change the states involved.
Now, next, Bob and Alice perform local measurements by testing
the properties
$$
P_b := |0 \rangle\langle 0|_b, \quad
P_a := |0 \rangle\langle 0|_a \, .
$$
If $\psi$ or $\psi'$ is the state of $\cH$, the states of
the local parts will be $(1/2) \1_a$ and $(1/2) \1_b$
respectively. The answer
is either YES or NO with equal property $1/2$ as seen from
$$
(|0 \rangle\langle 0|_a \otimes \1_b) \psi
= {1 \over \sqrt{2}} |00\rangle =
(\1_a \otimes |0 \rangle\langle 0|_b) \psi
$$
and from the similar relation with $\psi'$. There
is a strong correlation: Either both devices return YES or
both say NO. Therefore, if the input of Alice is $|0\rangle$,
the output is either YES for Alice as well as for Bob,
or it is NO for both.
If, however, $|1\rangle$ is the input of Alice, then
$|10\rangle$ becomes the state of $\cH$. It follows that
Alice gets necessarily NO and Bob YES.

We see that Bob and Alice would have the full information
of the message, Alice had encoded in her system, if both
parties could communicate their measurement results --
even if Alice has forgotten her original message.
No information is lost, but it is non-locally distributed
after testing the property $\psi$.

A particular interesting case is the transmission of information
from Alice to Bob, who knows neither the result of testing
the property $\psi$ nor has he obtained any information from
Alice. He knows, which property has been checked, but does not
know the result.

Though there is no classical information transfer,
Bob gets some information from Alice by testing in his
system property $P_b$. Considering all intermediate state
changes as done by a quantum black box, the process is
stepwise described by
$$
\{ |0 \rangle\langle 0|_a , |1 \rangle\langle 1|_a \}
\, \, \mapsto \, \,
\{ {1 \over 2} \1_b , |0 \rangle\langle 0|_b   \}
$$
and can be represented as an application of the stochastic cp-map
$$
\pmatrix{\omega_{00} & \omega_{01} \cr \omega_{10} & \omega_{11}}_a
\, \mapsto \,
{1 \over 2} \, \pmatrix{\omega_{00} + 2 \omega_{11} & 0 \cr 0
& \omega_{00}}_b \, .
$$
A message encoded by Alice with probabilities $q$ or $1-q$
per letters 0 or 1 carries an information
$H(q, 1-q)$. The Holevo bound for the quantum message
Bob obtains by measuring $P_b$ can be calculated to be
$$
H(1 - {1 \over 2} q , {1 \over 2} q) - q \, .
$$
Its maximum is reached at $q = (2/5)$. In that case Bob
receives approximately $0.322$ bit per letter, while Alice
has encoded her message with $0.962$ bit per letter.

What we have just discussed is a slight variation of
protocols invented independently by Aharonov and Albert,
\cite{AA81}, and by R.~D.~Sorkin \cite{So93}. The latter
claimed it to be an
example of a measurement ``forbidden by Einstein
causality''. More recently Beckman et al \cite{BGNP01},
adding an interesting collection of similar measurements,
have extended and sharpened Sorkin's assertion. On the
other hand, Vaidman \cite{Va01} presented teleportation
protocols of non-local measurements.
We, B.~Crell and me, \cite{CU},
think the causality considerations of Sorkin and
Beckmann et al not conclusive: While a measurement
allows for instantaneous changes of states, the output
of an apparatus includes classical information
processing which has to go on in the world region
the device is working. To detect the output of the
signal can only be possible in the intersection of
all future cones originating in  world points of
the measuring region. Bob can detect Alice's message
not before his world lines have crossed all the
future light cones originating from the world points
at which the measuring process is going on.
Hence, though the state change has taken
place, Bob can be informed only after a time delay of the
order ``radius of the measuring device / velocity
of light''.
Before that time has elapsed, the state change is hidden
to Bob -- as required by causality.

More accurate \cite{CU}, the rule with which quantum theory
outlines the defect of being not causal, is as follows.
Let $A$ and $B$ be two non-commuting observables which
we like to measure sequentially, say $A$ before $B$.
Let $G_A$ and $G_B$ denote the world region at which
the measurements should take place. Then {\em $G_B$
must be in the complete future of $G_A$,} that is
$G_B$ must be in the intersection of all forward
cones originating in the world points of $G_A$.
\medskip

The return to the general case of inverse EPR
with $\psi$ an arbitrary vector of a bi-partite
system with Hilbert space $\cH$ is formally
straightforward: Checking the property $\psi$ if the system
is in a product state $\phi_1^a \otimes \phi_1^b$ one comes across
$$
|\psi \rangle\langle \psi| \, \phi_1^a \otimes \phi_1^b =
\langle \psi , \phi_1^a \otimes \phi_1^b \rangle \, \psi \, .
$$
If Alice and Bob can communicate, and they can check with
which probability
 their states enjoy the property
 $\phi_2^a \otimes \phi_2^b$.
The transition amplitude for an affirmative answer
can be expressed, according to
(\ref{e6}),  by
$$
\langle \psi , \phi_1^a \otimes \phi_1^b \rangle \,
\langle \phi_2^a \otimes \phi_2^b , \psi \rangle =
\langle \phi_1^a , {\bf s}_{\psi}^{ab} \phi_1^b \rangle^*
\, \langle \phi_2^a , {\bf s}_{\psi}^{ab} \phi_2^b \rangle
\, .
$$

\section{Imperfect quantum teleportation}
Quantum teleportation has been invented by
Bennett et al \cite{BBCJPW93}. ``Perfect'' or
faithful quantum
teleportation starts within a product of
three Hilbert spaces of equal finite dimension and with a
maximal
entangled vector in the last two. It is triggered
by a von Neumann measurement in the first two spaces
using a basis of
maximally entangled vectors. The measurement
randomly chooses one of several quantum channels.
The information, which quantum channel has been activated,
is carried by the classical channel. It serves to reconstruct,
by a unitary move, the desired state at the destination.

All those possible ``perfect'' or ``tight'' schemes, together
with their dense coding counterparts, have been reviewed
by R.~F.~Werner \cite{We00}.

Following \cite{BBCJPW93} and analyzing their computations,
one can decompose the chosen
quantum channel into two parts, an inverse EPR and an
EPR setting. As one can identify  two particular s-maps
with them, one is tempted to use two general
s-maps. In doing so one can treat a more
general setup. But even in ``perfect'' circumstances
the explicit use of the mentioned decomposition may be
of some interest.

Let $\cH$ be a tri-partite Hilbert space
\begin{equation} \label{t1}
 \cH_{abc} = \cH_a \otimes \cH_b \otimes \cH_c \, .
\end{equation}
There is no restriction on the dimensions of the factor spaces.
The {\em input} is a vector $\phi^a \in \cH_a$, possibly
unknown, and a known vector $\varphi^{bc}$, the ``ancilla'',
out of $\cH_b \otimes \cH_c$. The
teleportation protocol is to start with the initial vector
\begin{equation} \label{t2}
\varphi^{abc} := \phi^a \otimes \varphi^{bc} \in \cH_{abc} \, .
\end{equation}
Now one performs a measurement on $\cH_a \otimes \cH_b$.
Instead of a complete von Neumann measurement we ask
just whether a property, given by a vector $\psi^{ab}$,
is present or not. In doing ``nothing'' on the c-system,
one is checking a local property of the
abc-system. If the check runs affirmative, the vector
state $\psi^{ab}$ is prepared in $\cH_{ab}$, inducing
a state change in the larger abc-system:
\begin{equation} \label{t3}
( |\psi^{ab} \rangle\langle \psi^{ab}| \otimes \1^c)
( \phi^a \otimes \varphi^{bc} )
= \psi^{ab} \otimes \phi^c,
\end{equation}
with a vector $\phi^c \in \cH_c$ yet to be determined.
Indeed,
$$
\phi^a \, \mapsto \, \phi^c
$$
represents the teleportation channel which is
triggered by an affirmative
check of the property defined by $\psi^{ab}$. Letting
$\phi^a$ as a free variable, we introduce the
{\em teleportation map} ${\bf t}^{ca}$ by
\begin{equation} \label{t4}
{\bf t}^{ca}  \phi^a \equiv {\bf t}^{ca}_{\psi, \varphi} \phi^a
= \phi^c, \quad
\psi \equiv \psi^{ab}, \, \, \varphi \equiv \varphi^{bc}
\end{equation}
The teleportation map ${\bf t}^{ca}$ is governed by
the {\em composition rule,}  \cite{Uh00a},
\begin{equation} \label{t5}
{\bf t}^{ca}_{\psi, \varphi} =
{\bf s}_{\varphi}^{cb} \, {\bf s}_{\psi}^{ba} \, .
\end{equation}
The s-maps being Hilbert-Schmidt, the t-maps must be
of trace class and linear. Indeed, {\em every trace class map
from $\cH_a$ into $\cH_c$ can be gained as a t-map, provided
its rank does not exceed the dimension of $\cH_b$.}
Of course, this fact can be obtained also directly, without
relying on the decomposition rule,
\cite{HHH96,MH99,AF00,TBL01,Ba01,RHFB01},
where also cases with a mixed ancilla have been studied.

\underline{Proof} of (\ref{t5}). \,
Let us abbreviate the left hand side of (\ref{t3})
by $\psi^{abc}$. Choosing in $\cH_b$
an ortho-normal basis $\{ \phi_j^b \}$ gives the
opportunity to write
$$
\varphi^{bc} =
\sum \phi_j^b \otimes
{\bf s}_{\varphi}^{cb} \phi_j^b
$$
and hence
$$
\psi^{abc} =
\psi^{ab}  \otimes \sum_j
\langle \psi^{ab} , \phi^a \otimes \phi_j^b \rangle \,
{\bf s}_{\varphi}^{cb} \phi_j^b \, .
$$
We choose in $\cH_a$ an ortho-normal basis,
$\{ \phi_k^a \}$,
to resolve the scalar product in the last equation:
$$
\psi^{abc} =
\psi^{ab} \otimes \sum_{jk}
\langle \phi_k^a , \phi^a \rangle \, \langle
{\bf s}_{\psi}^{ba}  \phi_k^a , \phi_j^b \rangle \,
{\bf s}_{\varphi}^{cb} \phi_j^b \, .
$$
Using anti-linearity,
$$
\psi^{abc} =
\psi^{ab} \otimes {\bf s}_{\varphi}^{cb}  \sum_k
\langle \phi^a , \phi_k^a \rangle \, \sum_j
\langle \phi_j^b , {\bf s}_{\psi}^{ba}
\phi_k^a \rangle \, \phi_j^b
$$
The summation over $j$ results in
${\bf s}_{\psi}^{ba}  \phi_k^a$.
Next, again by anti-linearity, the sum over $k$ comes down to
$$
{\bf s}_{\psi}^{ba} \sum_k \langle \phi_k^a ,
\phi^a \rangle \, \phi_k^a =
{\bf s}_{\psi}^{ba} \phi^a
$$
and we get finally
$$
\psi^{abc} = \psi^{ab} \otimes
{\bf s}_{\varphi}^{cb} \, {\bf s}_{\psi}^{ba}
 \phi^a
$$
and the composition rule is proved.
\medskip

\underline{Distributed measurements}

The next aim is to present an extension of the composition
rule to multi-partite systems.
In a multi-partite system one can distribute the measurements
and the entanglement resources over some pairs of subsystems.
With an odd number of subsystems we get
{\em distributed teleportation,} with an even number something like
{\em distributed EPR.}

At first let us see, as an example, distributed teleportation
with five subsystems.
\begin{equation} \label{dt1}
\cH = \cH_a \otimes \cH_b \otimes \cH_c \otimes
\cH_d \otimes \cH_e \, .
\end{equation}
The input is an unknown vector $\phi^a \in \cH_a$, the
ancillarian vectors are selected from the $bc$- and the
$de$-system,
\begin{equation} \label{dt2}
\varphi^{bc} \in \cH_{bc} = \cH_b \otimes \cH_c, \quad
\varphi^{de} \in \cH_{de} = \cH_d \otimes \cH_e,
\end{equation}
and the vector of the total system we are starting with is
\begin{equation} \label{dt3}
 \varphi^{abcde} = \phi^a \otimes
\varphi^{bc} \otimes \varphi^{de} \, .
\end{equation}
The channel is triggered by measurements in the $ab$- and
in the $cd$-system.
Suppose these measurements are successful and they
 prepare the vector states
\begin{equation} \label{dt4}
\psi^{ab} \in \cH_{ab} = \cH_a \otimes \cH_b, \quad
\psi^{cd} \in \cH_{cd} = \cH_c \otimes \cH_d \, .
\end{equation}
Then we get the relation
\begin{equation} \label{dt5}
( |\psi^{ab} \rangle\langle \psi^{ab}| \otimes
 |\psi^{cd} \rangle\langle \psi^{cd}| \otimes \1^e )  \varphi^{abcde}
= \psi^{ab} \otimes \psi^{cd} \otimes \phi^e
\end{equation}
and the vector $\phi^a$ is mapped onto $\phi^e$.
Introducing the s-maps corresponding to the vectors
$$
\psi^{ab} \to {\bf s}^{ba}, \quad
\varphi^{bc} \to {\bf s}^{cb}, \quad
\psi^{cd} \to {\bf s}^{dc}, \dots,
$$
the {\em factorization rule} becomes
\begin{equation} \label{dt6}
\phi^e = {\bf t}^{ea} \phi^a, \quad
{\bf t}^{ea} = {\bf s}^{ed} \, {\bf s}^{dc} \,
{\bf s}^{cb} \, {\bf s}^{ba} \, .
\end{equation}

Next we consider a setting with {\em four} Hilbert
spaces, $\cH_b$ to $\cH_e$. The input state is
$$
\varphi^{bcde} = \varphi^{bc} \otimes \varphi^{de}
$$
and we perform a test to check whether the property
$\psi^{cd}$ is present or not.
Let the answer be YES. Then the subsystems bc and
de become disentangled. The cd system gets $\psi^{cd}$
and, hence, the entanglement of this vector state.
The previously unentangled systems $\cH_b$ and $\cH_e$
will now be entangled.

The newly prepared state is
\begin{equation} \label{dt7}
\chi^{bcde} :=
(\1_b \otimes |\psi^{cd} \rangle\langle \psi^{cd}|
\otimes \1_e ) \, \varphi^{bcde} \, .
\end{equation}
With
$$
\psi^{cd} = \sum \lambda_j \phi^c_j \otimes \phi^d_j
$$
we obtain
$$
\chi^{bcde} = \sum \lambda_j \lambda_k
[(\1_b \otimes |\phi^c_j \rangle\langle \phi^c_j |) \varphi^{bc}]
\otimes
[( |\phi^d_j \rangle\langle \phi^d_j | \otimes \1_e) \varphi^{de}]
\, .
$$
Let us denote just by ${\bf s}^{bc}$ and ${\bf s}^{de}$
the s-maps of $\varphi^{bc}$ and $\varphi^{de}$
respectively. They allow to rewrite $\chi^{bcde}$ as
$$
\chi^{bcde} =
\sum \lambda_j \lambda_k ({\bf s}^{bc} \phi^c_k \otimes \phi^c_j)
\otimes
(\phi^d_j \otimes {\bf s}^{ed} \phi^d_k)
$$
which is equal to
\begin{equation} \label{dt8}
\chi^{bcde} =
\sum \lambda_k ({\bf s}^{bc} \phi^c_k) \otimes \psi^{cd}
\otimes ({\bf s}^{ed} \phi^c_d) \, .
\end{equation}
The Hilbert space $\cH_c \otimes \cH_d$ is decoupled
from $\cH_b$ and $\cH_e$. The vector state of the latter
can be characterized by a map from $\cH_c \otimes \cH_d$
into $\cH_b \otimes \cH_e$.
\begin{equation} \label{dt9}
\varphi^{be} := ( {\bf s}^{bc} \otimes {\bf s}^{ed} ) \,
\psi^{cd}
\end{equation}
is indicating how the entanglement within the be-system
is arising, and how the  three vectors involved come together
to achieve it.
\medskip

\underline{Addendum:} \, A rearrangement lemma.

The starting point is a collection of bi-partite
spaces and vectors,
\begin{equation} \label{rea1}
\psi_j \in \cH_{ab}^j , \quad \cH_{ab}^j = \cH_a^j \otimes \cH_b^j,
\quad j = 1, \dots, m
\end{equation}
from which we build
\begin{equation} \label{rea2}
\cH_{ab} = \cH_{ab}^1 \otimes \dots \otimes \cH_{ab}^j,
\quad
\psi = \psi_1 \otimes \dots \otimes \psi_m \, .
\end{equation}
We abbreviate the s-maps accordingly,
\begin{equation} \label{rea3}
\psi_j \, \leftrightarrow \, {\bf s}^{ab}_j
\, \leftrightarrow \,    {\bf s}^{ba}_j
\end{equation}
We now change to the rearranged Hilbert space
\begin{equation} \label{rea4}
\cH_{AB} = \cH_A \otimes \cH_B =
(\cH_a^1 \otimes \dots \cH_a^m) \, \otimes \,
\cH_b^1 \otimes \dots \cH_b^m) \, .
\end{equation}
The Hilbert spaces (\ref{rea2}) and (\ref{rea4}) are unitarily
equivalent in a canonical way:
\begin{equation} \label{rea5}
V \, : \quad \cH_{ab} \, \mapsto \, \cH_{AB}
\end{equation}
is defined to be the linear map satisfying
\begin{equation} \label{rea6}
V \, ( \phi^a_1 \otimes \phi^b_1 \otimes \dots \otimes
\phi^a_m \otimes \phi^b_m ) =
( \phi^a_1 \otimes \dots \otimes \phi^a_m ) \, \otimes \,
( \phi^b_1 \otimes \dots \otimes \phi^b_m )
\end{equation}
This is a unitary map, $V^{-1} = V^*$.

Assume we need the s-maps of
\begin{equation} \label{rea7}
\varphi := V \, \psi
\end{equation}
with $\psi$ given by (\ref{rea2}). The rearrangement lemma
we have in mind reads
\begin{equation} \label{rea8}
{\bf s}^{AB}_{\varphi} = V \,
( {\bf s}^{ab}_1 \otimes \dots \otimes {\bf s}^{ab}_m ) \, V^{-1}
\, .
\end{equation}
The proof uses the fact that both sides are multi-linear
in the vectors $\psi_j$. Therefore, it suffices to establish
the assertion in the case, the $\psi_j$ are product vectors.
But then the proof consists of some lengthy but easy to
handle identities.

\section{Polar decompositions}
Let us come back to the s-maps. It is worthwhile to study
their polar decompositions. As we already know (\ref{e8})
it is evident that we should have
\begin{equation} \label{p1}
{\bf s}_{\psi}^{ba} = (\varrho_{\psi}^b)^{1/2} {\bf j}_{\psi}^{ba}
= {\bf j}_{\psi}^{ba} (\varrho_{\psi}^a)^{1/2},
\end{equation}
$$
{\bf s}_{\psi}^{ab} = (\varrho_{\psi}^a)^{1/2} {\bf j}_{\psi}^{ab}
= {\bf j}_{\psi}^{ab} (\varrho_{\psi}^b)^{1/2} \, .
$$
The j-maps are anti-linear partial isometries with left (right)
supports equal to the support of their left (right) positive
factor. From Alice's point of view, who can know her
reduced density operator but not the state from which it
is reduced, ${\bf j}_{\psi}^{ab}$ is a
non-commutative phase. It is in discussion whether
and how relative phases of this kind can be detected
experimentally.

One outcome of the polar decomposition is a unique labelling
of purifications. If $\varrho^a$ denotes a density operator on
$\cH_a$, then all its purifications can be gained by the chain
$$
\varrho^a \mapsto {\bf j}^{ba} (\varrho^a)^{1/2} =
{\bf s}_{\psi}^{ba} \mapsto \psi
$$
where ${\bf j}^{ba}$ runs through all those anti-linear isometries
from a to b whose right supports are equal to the support of
$\varrho^a$.

The uniqueness of the polar decomposition and
(\ref{e8}) yields
\begin{equation} \label{p2}
( {\bf j}_{\psi}^{ba} )^* = {\bf j}_{\psi}^{ab}, \quad
\varrho_{\psi}^b =
{\bf j}_{\psi}^{ba} \, \varrho_{\psi}^a \, {\bf j}_{\psi}^{ab}
\, .
\end{equation}
Now we can relate the expectation values of the reduced density
operators: Assume the bounded operators $A$ and $B$ on
$\cH_a$ and $\cH_b$ are such that
\begin{equation} \label{p3}
B^* \, {\bf j}_{\psi}^{ba} = {\bf j}_{\psi}^{ba} A \, .
\end{equation}
Then one gets, as  a little exercise in anti-linearity,
\begin{equation} \label{p4}
\T \, \varrho_{\psi}^a A = \T \, \varrho_{\psi}^b B \, .
\end{equation}
It is possible to
express the condition (\ref{p3}) for the validity of (\ref{p4})
by an anti-linear operator $J_{\psi}$ acting on
$\cH_a \otimes \cH_b$. To this end we define $J_{\psi}$
as the anti-linear extension of
\begin{equation} \label{p5}
J_{\psi} ( \phi^a \otimes \phi^b ) =
{\bf j}_{\psi}^{ab} \phi^b \otimes {\bf j}_{\psi}^{ba} \phi^a
\, .
\end{equation}
With this definition it is to be seen that (\ref{p3})
is as strong as
\begin{equation} \label{p6}
J_{\psi} (A \otimes B) = (A \otimes B)^* J_{\psi} \, .
\end{equation}
(\ref{p5}) is a crossed tensor product, $\tilde \otimes$.
With every
pair of maps, one from $\cH_a$ to $\cH_b$ and one
in the opposite direction, and both either linear or
anti-linear, one can build the crossed tensor product
$\tilde \otimes$. An important example is (\ref{p5}),
where the two factors are
j-maps. We may formally write
$$
J_{\psi} = {\bf j}_{\psi}^{ab} \,
\tilde \otimes  \, {\bf j}_{\psi}^{ba}
$$
for the just defined anti-linear operator acting on $\cH_{ab}$.

Now let the factors of $\cH_{ab}$ be of equal dimension and
$\psi$ ``completely entangled''.
In a more mathematical language $\psi$ is called a cyclic
and separating
vector, a so-called GNS-vector\footnote{GNS stands for
I.~M.~Gelfand, M.~A.~Naimark, I.~E.~Segal}
or a ``GNS vacuum'', for the representation
$$
A \, \mapsto \, A \otimes \1_b
$$
of the algebra $\cB(\cH_a)$. In this context,
$J_{\psi}$ is an elementary example of Tomita-Takeski's
{\em modular conjugation.}
That $\psi$ is completely entangled can be expressed also in terms
of s-maps: ${\bf s}_{\psi}^{ab}$ must be invertible. (Its
inverse, if it exists, must be unbounded for infinite
dimensional Hilbert spaces.)

There are two further operators,
particulary tied to the modular conjugation.
The first is introduced by
\begin{equation} \label{p7}
(A \otimes \1_b) \, \psi = S_{\psi} (A^* \otimes \1_b) \psi \, .
\end{equation}
$S_{\psi}$ can also be gained by the help of the twisted
cross product
\begin{equation} \label{p8}
S_{\psi} = ({\bf s}_{\psi}^{ba})^{-1} \, \tilde \otimes \,
{\bf s}_{\psi}^{ba} \, .
\end{equation}
It is standard to write the polar decomposition
of the anti-linear S-operator
\begin{equation} \label{p9}
S_{\psi} = J_{\psi} \sqrt{ \Delta_{\psi}} \, .
\end{equation}
$\Delta_{\psi}$ is called the Tomita-Takesaki
{\em modular operator.} The distinguished role of these
and similar ``modular objects'' becomes apparent in
the theory of general von Neumann algebras where they play
an exposed and quite natural role. From them I borrowed the
notations for the s- and the j-maps.
In the elementary case we are dealing with, one has
$$
\Delta_{\psi} = \varrho_{\psi}^a \otimes
(\varrho_{\psi}^b)^{-1} \, .
$$
See \cite{Haag93} for a physically motivated introduction.
Further relations between the s- and j-maps and to modular
objects can be found in \cite{Uh00a} and \cite{Uh00b}.

\section{From vectors to states}
With $\varrho \equiv \varrho^{ab}$ we may write similar to
(\ref{e4}),
\begin{equation} \label{st1}
(|\phi^a \rangle\langle \phi^a| \otimes \1_b) \, \varrho^{ab}
(|\phi^a \rangle\langle \phi^a| \otimes \1_b) =
|\phi^a \rangle\langle \phi^a| \otimes
\Phi^{ba}_{\varrho}(|\phi^a \rangle\langle \phi^a|),
\quad
\forall \phi^a \in \cH_a
\end{equation}
For every decomposition
\begin{equation} \label{st2}
\varrho^{ab} = \sum c_{jk} |\psi_j \rangle\langle \psi_k|,
\quad \cH_{ab} \ni \psi_j \, \leftrightarrow \, {\bf s}_j^{ab}
\end{equation}
there is a representation
\begin{equation} \label{st3}
\Phi^{ba}_{\varrho}(|\phi^a \rangle\langle \phi^a|) =
\sum c_{jk} {\bf s}_j^{ba} |\phi^a \rangle\langle \phi^a|)
 {\bf s}_k^{ab} \, .
\end{equation}
Similarly one defines $\Phi_{\varrho}^{ab}$.  The maps are linear
in $\varrho^{ab}$ and can be defined for every trace class
operator $\varrho$. Moreover, their domain of definition can
be extended to the bounded operators of the subsystems: Let
$X$ and $Y$ denote bounded operators on $\cH_a$ and $\cH_b$
respectively, then
\begin{equation} \label{st4}
X \mapsto \Phi^{ba}_{\varrho}(X),
\quad
Y \mapsto \Phi^{ab}_{\varrho}(Y)
\end{equation}
are well defined and anti-linear in $X$ or $Y$. The equation
\begin{equation} \label{st5}
\T \, X \Phi^{ab}_{\varrho}(Y^*) =
\T \, Y \Phi^{ba}_{\varrho}(X^*) =
\T \, \varrho \, (X \otimes Y)
\end{equation}
is valid. Proving them at first for finite linear combinations of
rank one operators, one finds the maps (\ref{st4}) mapping the
bounded operators of one subsystem into the trace class operators
of the other one. Indeed, the finite version of (\ref{st5})
provides us with estimates like
\begin{equation} \label{st5a}
\parallel \Phi^{ba}_{\varrho}(X^*) \parallel_1 \, \leq \,
\parallel X \parallel_{\infty} \,
\parallel \varrho \parallel_1 \, .
\end{equation}
We now have a one-to one correspondence
\begin{equation} \label{st5b}
\Phi^{ab}_{\varrho} \, \leftrightarrow \, \Phi^{ba}_{\varrho}
\, \leftrightarrow \, \varrho
\end{equation}
That we have a map from the bounded operators of $\cH_a$ into
the trace class operators of $\cH_b$ is physically quite nice.
It is an opportunity to reflect on testing a property $P_a$
of $\cH_a$ once more, but under the condition that
$\varrho \equiv \varrho^{ab}$ is in any
(normal) state. The rank of $P_a$ is not necessarily
finite. The rule of L\"uders, \cite{Lued51}, says that
the prepared state is $\omega^a := P_a \varrho^a P_a$ if
one finds the property $P_a$ valid
and $\varrho^a$ is the reduced density matrix
of $\varrho$ in the a-system before the test.
The EPR channel asks for $\omega^b$,
the
density operator of the b-system after an affirmative checking
of the property $P_a$. This density operator is given by a
$\Phi$-map:

{\em If $\varrho$ is the density operator of $\cH_{ab}$ and if
a local measurement establishes property $P_a$, then the state
$\omega^b$ of the b-system is given by}
\begin{equation} \label{st6}
\omega^b = \Phi_{\varrho}^{ba}(P_a) \, .
\end{equation}
The proof is by looking at the effect in the bi-partite system
resulting from a local measurement. Let $\phi^a_j$ be
a basis of the support space of $P_a$. One obtains
$$
(P_a \otimes \1_b) \, |\psi \rangle\langle \psi| \, (P_a \otimes \1_b) =
\sum |\phi_j^a \rangle\langle \phi_k^b| \otimes
{\bf s}_{\psi}^{ba} |\phi^a_k \rangle\langle \phi^a_j| \,
{\bf s}_{\psi}^{ab}
$$
and this is, up to normalization, the state prepared by the local
measurement. Next we sandwich the equation between $\1_a \otimes B$
and take the trace. At the left hand we
get $\T \, \omega^b B$. On the right
we obtain $\Phi_{\varrho}^{ba}(P_a)$.
Now we have seen from (\ref{st5}) that (\ref{st6})
is correct for pure states.
By linearity and (\ref{st5a}) we get the assertion.

It may be worthwhile to compare (\ref{st5}) with the now well known
``duality'' between super-operators $T$ of $\cH_a$ and operators
on $\cH_a \otimes \cH_b$. Here the Hilbert spaces are of equal
finite dimension.
One selects a maximally entangled vector $\psi$ and
defines
\begin{equation} \label{st7}
\rho :=
( T \otimes {\rm id}_b )(|\psi \rangle\langle \psi|)
\end{equation}
to express the structure of $T$ by that
of $\rho$. This trick is due to A.~Jamiolkowski,
\cite{Jamiol72}, and is now refined and much in use after the
papers of B.~Terhal \cite{Terhal98} and of Horodecki et al
\cite{Horo98}. Comparing (\ref{st3}) and (\ref{st4}), one can
connect both approaches as follows:

From $\varrho$ we get
a map $\Phi_{\varrho}^{ab}$. From a maximally entangled $\psi$
we get an anti-linear map ${\bf s}_{\psi}^{ab}$, enabling the
correspondence (\ref{st7}) to be expressed by
\begin{equation} \label{st8}
\varrho \, \leftrightarrow \, T , \quad T(X) =
{\bf s}_{\psi}^{ab} \,
\Phi_{\varrho}^{ba}(X) \, {\bf s}_{\psi}^{ba}
\end{equation}
 In a certain way, anti-linearity
is the prize for eliminating the reference state $\psi$ in
Jamiolkowski's
approach.

I thank Bernd Crell for valuable comments.

email: armin.uhlmann@itp.uni-leipzig.de

\begin{thebibliography}{**}
\bibitem{CU} For the discussion of that point of view
I am indebted to B.~Crell, Leipzig.
\end{thebibliography}
\end{document}